\documentclass[twocolumn,showpacs,prb,amsmath,amssymb]{revtex4}

\usepackage{graphicx}
\usepackage{dcolumn}
\usepackage{bm}

\begin{document}
\draft

\title{
A theoretical analysis on 
highly spin-polarized transport of iron nitride Fe$_4$N 
 }
\author{
Satoshi~Kokado$^{1}$
\footnote{Electronic mail:tskokad@ipc.shizuoka.ac.jp}, 
Nobuhisa~Fujima$^1$, 
Kikuo~Harigaya$^2$, 
Hisashi~Shimizu$^3$, and 
Akimasa~Sakuma$^4$
}
\address{
$^1$Faculty of Engineering, Shizuoka University, Hamamatsu 432-8561, Japan \\
$^2$Nanotechnology Research Institute, AIST, Tsukuba 305-8568, Japan \\
$^3$Faculty of Science, Shinshu University, Matsumoto 390-8621, Japan \\
$^4$Graduate School of Engineering, Tohoku University, Sendai 980-8579, Japan
}
\date{\today}


\begin{abstract}
In order to propose a ferromagnet exhibiting highly spin-polarized transport, 
we theoretically analyzed the spin polarization ratio 
of the conductivity of the bulk Fe$_4$N 
with a perovskite type structure, 
in which N is located at the body center position of fcc-Fe. 
The spin polarization ratio is defined by 
$P = ( \sigma_\uparrow - \sigma_\downarrow ) 
/ ( \sigma_\uparrow + \sigma_\downarrow )$, 
with $\sigma_{\uparrow(\downarrow)}$ being the conductivity 
at zero temperature of the up spin (down spin). 
The conductivity is obtained by using the Kubo formula 
and the Slater-Koster tight binding model, 
where parameters are determined 
from the least-square fitting of the dispersion curves 
by the tight binding model 
to those by the first principles calculation. 
In the vicinity of the Fermi energy, 
$|P|$ takes almost 1.0, indicating perfectly spin-polarized transport. 
In addition, 
by comparing Fe$_4$N to fcc-Fe (Fe$_4$N$_0$) in the ferromagnetic state 
with the equilibrium lattice constant of Fe$_4$N, 
it is shown that the non-magnetic atom 
N plays an important role in increasing $|P|$. 
\end{abstract}

\pacs{72.25.Ba}

\maketitle

Recently, highly efficient spin-electronics devices 
operating at room temperature 
have been extensively developed 
for applications to the magnetic memory and the magnetic sensor. 
A typical device has 
ferromagnetic tunnel junctions consisting of 
a ferromagnetic electrode (FME)/insulator/FME, 
which exhibits 
a large magnetoresistance (MR) 
effect.~\cite{Inomata,Sakuraba,Yuasa1,Yuasa2,Hayakawa1,Hayakawa2,Kokado} 
The efficiency of the MR effect is often defined by 
${\rm the~MR~ratio} = (R_{\mbox{\tiny AP}} - R_{\mbox{\tiny P}})/
R_{\mbox{\tiny AP}}$, 
with $R_{\mbox{\tiny P}}$ and $R_{\mbox{\tiny AP}}$ 
being the resistance of the parallel and anti-parallel 
magnetization configurations of FMEs, respectively. 
Experimentally, 
regarding junctions with an electrode of a half-metallic ferromagnet, 
Co$_2$Cr$_{0.6}$Fe$_{0.4}$Al/Al-O/CoFe/NiFe/IrMn/Ta junctions 
exhibited an MR ratio of 16 \% at room temperature,~\cite{Inomata} 
and Co$_2$MnSi/Al-O/Co$_{75}$Fe$_{25}$ 
had an MR ratio of 70 \% at room temperature.~\cite{Sakuraba} 
Regarding junctions of electrodes with usual ferromagnets, 
single-crystal Fe(001)/MgO(001)/Fe(001) 
junctions exhibited MR ratios 
of 88 \%~\cite{Yuasa1} and 180 \%~\cite{Yuasa2} at room temperature. 
Furthermore, CoFeB/MgO/CoFeB junctions achieved MR ratios 
of 260 \%~\cite{Hayakawa1} and 355 \%~\cite{Hayakawa2} 
(the world's highest value) at room temperature, 
although the crystal structure of CoFeB 
and the role of the light element B on the spin-polarized transport 
have not been clarified yet. 
Generally, the MR ratio becomes large 
with increasing the spin polarization 
of the conduction electron in the FME. 
In the future, an FME with more highly spin-polarized electrons, 
which would result in larger MR ratios, 
will be strongly desired 
from the viewpoint of the development of highly efficient MR devices.

Towards proposal of such an FME, 
we extracted an idea to obtain electrodes 
exhibiting the highly spin-polarized transport at room temperature. 
We found that ferromagnets 
consisting of $magnetic~elements$ and $light~elements$, such as CoFeB, 
might be very useful as the electrodes.

We consider Fe as a representative magnetic element in this idea. 
We define the spin polarization (SP) ratio 
about the density of states (DOS) of the bulk system, 
$P_{\mbox{\tiny DOS}}(E) $, as 
\begin{eqnarray}
P_{\mbox{\tiny DOS}}(E) = \frac{ D_\uparrow (E) - D_\downarrow (E)}
{D_\uparrow (E)+ D_\downarrow(E)}, 
\end{eqnarray}
with $D_s (E)$ being the DOS of spin $s$ (=$\uparrow$ or $\downarrow$) 
for the energy $E$. 
Then, $|P_{\mbox{\tiny DOS}} (E_{\mbox{\tiny F}})|$ 
at the Fermi energy $E_{\mbox{\tiny F}}$ 
of fcc-Fe in the ferromagnetic state 
is evaluated to be about 0.7 [see Fig. 2(c)], 
which is about 2.3 times as large as that of bcc-Fe,~\cite{bcc-Fe} 
where the most stable state of bcc-Fe (fcc-Fe) is 
ferromagnetic (not ferromagnetic~\cite{Hoshino}). 
This may indicate that 
the highly spin-polarized transport is realized 
for materials closely related to fcc-Fe. 
The conductivity, including the velocity of electrons, 
will give an answer to the realizability of such transport.

We focus on a ferromagnet containing fcc-Fe and a light element, 
Fe$_4$N with a perovskite-type 
structure,~\cite{Jack,Shirane,Nagakura,Zhou,Sakuma,Ishida} 
in which N is located at the body center position of fcc-Fe. 
This ferromagnet has a Curie temperature of 761 K.~\cite{Nagakura} 
It should be noted that 
studies on the spin-polarized transport of Fe$_4$N 
have scarcely been performed so far, 
although other properties have been 
experimentally~\cite{Jack,Shirane,Nagakura} 
and theoretically~\cite{Zhou,Sakuma,Ishida} investigated. 
We are particularly interested in 
not only the SP ratio on the transport of Fe$_4$N 
but also the role of the light (or non-magnetic) element N 
on the transport.

In this paper, we analyzed the SP ratio of conductivity of Fe$_4$N 
in order to elucidate the spin-polarized transport. 
The conductivity 
was obtained for each spin and each orbital of the bulk system 
using the first principles (FP) calculation 
and the tight binding (TB) model calculation. 
Consequently, we found that 
Fe$_4$N exhibits an extremely highly spin-polarized transport 
and that N plays an important role 
in the transport.

A calculation method is introduced 
to obtain the conductivity and the SP ratio. 
The method is a combination of 
(i) the FP calculation,~\cite{Kresse} 
(ii) the TB model,~\cite{Slater} 
and (iii) the Kubo formula.~\cite{Kubo} 
Details of each are given below:

\begin{itemize}
\item[(i)] The FP calculations are performed 
by the Vienna Ab-initio Simulation Package (VASP) code~\cite{Kresse} 
based on the spin-polarized density functional theory, 
where we employ the generalized gradient approximation of 
Perdew and Wang (PW91)~\cite{Perdew} and 
ultrasoft pseudopotentials 
to describe the core electrons. 
The cutoff energy for the plane wave basis is 
237.51 eV for Fe and 348.10 eV for Fe$_4$N, 
and the Monkhorst-Pack set~\cite{Monkhorst} 
of 8$\times$8$\times$8 ${\bf k}$ points is used.
\item[(ii)]	The Slater-Koster TB model~\cite{Slater} is used 
with taking into account the 3d, 4s, and 4p orbitals for Fe 
and the 2s and 2p orbitals for N as well as 
interactions up to the third-nearest neighbor atoms. 
The Slater-Koster parameters 
of potential energies and transfer integrals~\cite{Slater} 
are here determined 
from the least-square fitting of the dispersion curves by the TB model 
to those by the FP calculation 
at the equilibrium lattice constant. 
The fitting is done in energy regions 
from the lowest energy to $E_{\mbox{\tiny F}}$ + 5 eV 
for bcc-Fe and fcc-Fe and 
from the lowest energy to $E_{\mbox{\tiny F}}$ + 2 eV for Fe$_4$N. 
The number of parameters 
is 34 for bcc-Fe and fcc-Fe and 87 for Fe$_4$N. 
\item[(iii)] The Kubo formula~\cite{Kubo} 
and the Slater-Koster TB model with the determined parameters are 
used to calculate the conductivity at zero temperature. 
In this calculation, we utilize the theory given 
by Tsymbal {\it et al}.~\cite{Tsymbal} 
The total conductivity of spin $s$ (=$\uparrow$ or $\downarrow$), 
$\sigma_s (E)$, is written by, 
\begin{eqnarray}
\sigma_s (E) = \displaystyle{\sum_i} \sigma_{i,s} (E),  
\end{eqnarray}
where $\sigma_{i,s}(E)$ is the conductivity 
of orbital $i$ (=4s, 4p, 3d orbitals, and so on) and spin $s$. 
This $\sigma_{i,s}(E)$ includes 
the velocity of electrons and the Green's function. 
The Green's function has a single parameter $\gamma$ 
in the second-order self energy due to the weak electron-impurity interaction, 
where $\gamma$ characterizes the degree of electron-impurity scattering. 
In detail, $\gamma$ is related to 
the lifetime of the electron of orbital $i$ and spin $s$, 
$\tau_{i,s}$, via $1/\tau_{i,s} = (2 \pi/\hbar) \gamma^2 D_{i,s} (E)$, 
with $D_{i,s} (E)$ being 
the partial DOS for $i$ and $s$.~\cite{Tsymbal} 
At present, $\gamma$ is set to be 0.5 eV, which was previously chosen 
so as to reproduce the resistivity of copper.~\cite{Tsymbal} 
We also use the diagonal approximation 
for the self energy, 
in which 
the scattering of electrons due to impurities 
is allowed for the same energy levels 
but forbidden 
between different energy levels.~\cite{Tsymbal} 
The SP ratio is then defined by 
\begin{eqnarray}
P(E)= \frac{\sigma_\uparrow (E) - \sigma_\downarrow (E) }
{\sigma_\uparrow (E) + \sigma_\downarrow (E)}.
\end{eqnarray}
\end{itemize}

We first compare the results of bcc-Fe 
by the present method with the previous ones. 
The equilibrium lattice constant obtained 
by using the FP calculation 
is estimated to be 2.84 \AA~
with an error of about 1 \% for the experimental value 
of 2.87 \AA.~\cite{Kittel}
The total conductivity at $E_{\mbox{\tiny F}}$ of the down spin 
$\sigma_\downarrow (E_{\mbox{\tiny F}})$ is larger than 
$\sigma_\uparrow (E_{\mbox{\tiny F}})$. 
The SP ratio at $E_{\mbox{\tiny F}}$, $P(E_{\mbox{\tiny F}})$, therefore 
takes a negative value, $-$0.20, 
and it qualitatively agrees with the previous result of 
about $-$0.26,~\cite{Tsymbal} 
which was obtained by a similar method 
combined with the FP calculation 
within the local density approximation.

In the following, we investigate 
the equilibrium lattice constant, $D_s(E)$, and $P_{\mbox{\tiny DOS}}(E)$ 
for Fe$_4$N using the FP calculation. 
The equilibrium lattice constant is evaluated as 3.810 \AA, 
which has an error of less than 1 \% 
for an experimental value of 3.795 \AA.~\cite{Nagakura} 
As shown in Figs. 1(a) and 1(c), 
$D_\downarrow (E_{\mbox{\tiny F}})$ is higher than 
$D_\uparrow (E_{\mbox{\tiny F}})$. 
Partial DOSs~\cite{Sakuma} 
schematically illustrated in Fig. 1(b) 
show that 3d orbitals are dominant around $E_{\mbox{\tiny F}}$, 
and 4s and 4p (4s-4p) orbitals are mainly located 
in an energy region higher than $E_{\mbox{\tiny F}}$, 
while each orbital of N atoms mostly exists 
in an energy region lower than $E_{\mbox{\tiny F}}$. 
Furthermore, 
$P_{\mbox{\tiny DOS}}(E_{\mbox{\tiny F}})$ obtained by using 
$D_{s}(E_{\mbox{\tiny F}})$ of the FP calculation 
is evaluated to be $-$0.6 [see Fig. 1(c)].

Using the TB model 
with the parameters determined from the fitting of the dispersion curves, 
we obtain $D_s (E)$ and $P_{\mbox{\tiny DOS}}(E)$ 
for $-1~{\rm eV} \le E \le 1~{\rm eV}$. 
As seen from Fig. 1(c), 
$D_s (E_{\mbox{\tiny F}})$ and $P_{\mbox{\tiny DOS}}(E_{\mbox{\tiny F}})$ 
of the TB model 
agree well with the respective ones of the FP calculation. 

With the use of the TB model and the Kubo formula 
we calculate $\sigma_{i,s} (E)$, $\sigma_s (E)$, and $P(E)$ 
for $-1~{\rm eV} \le E - E_{\mbox{\tiny F}} \le 1~{\rm eV}$. 
The results are shown in Fig. 1(d). 
For $E \le E_{\mbox{\tiny F}}$, 
$\sigma_{\mbox{\tiny 3d},\uparrow}(E)$ 
becomes relatively large owing to 
the high DOS of the 3d orbitals of the up spin. 
For $E - E_{\mbox{\tiny F}} \ge 0.6~{\rm eV}$, 
the 4s-4p orbitals of the up spin contribute strongly to 
$\sigma_\uparrow (E)$ in spite of their low DOSs 
because their orbitals have large velocities. 
For $0~{\rm eV} < E - E_{\mbox{\tiny F}} < 0.6~{\rm eV}$, 
each $\sigma_{i,\uparrow} (E)$ has a pronounced valley 
reflecting the low DOS of the up spin. 
In this energy region, 
although the DOS of the up spin is actually lower than 
that of the FP calculation, 
the qualitative behavior of $\sigma_{i,\uparrow} (E)$ 
appears to be valid 
because the DOS of the FP calculation 
has very few components of the 4s-4p orbitals 
and it is low. 
On the other hand, each $\sigma_{i,\downarrow} (E)$ 
is almost flat, and the 3d orbitals of the down spin contribute largely to 
$\sigma_\downarrow (E)$ 
because of their high DOS. 
The total conductivity at $E_{\mbox{\tiny F}}$ of the down spin 
$\sigma_\downarrow (E_{\mbox{\tiny F}})$ is much larger than 
$\sigma_\uparrow (E_{\mbox{\tiny F}})$. 
The SP ratio $P(E_{\mbox{\tiny F}})$ therefore takes almost $-$1.0, 
indicating perfectly spin-polarized transport. 
The magnitude of the SP ratio 
$|P(E_{\mbox{\tiny F}})|$ is about 5.0 times as large as that of bcc-Fe.

\begin{figure}[ht]
\begin{center}
\includegraphics[width=0.73\linewidth]{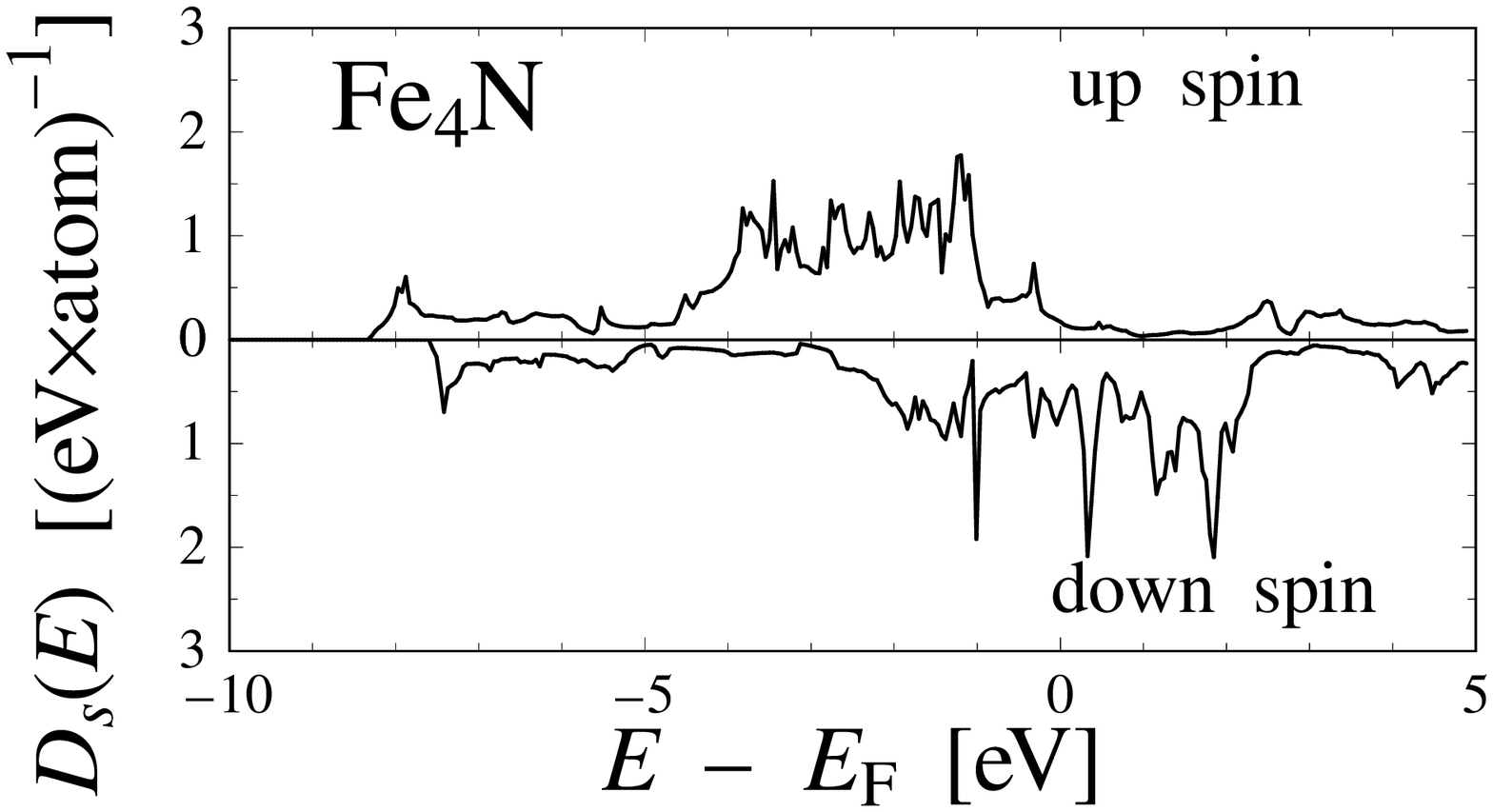}\\[-0.9cm]
\hspace*{-7.5cm}(a) \\
\vspace{0.6cm}
\hspace*{0.5cm}
\includegraphics[width=0.57\linewidth]{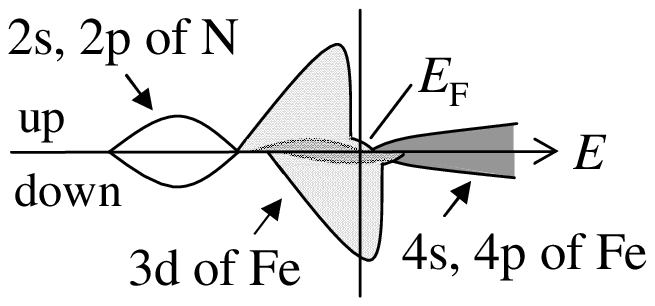}\\[-0.9cm]
\hspace*{-7.5cm}(b) \\
\vspace{0.22cm}
\includegraphics[width=0.73\linewidth]{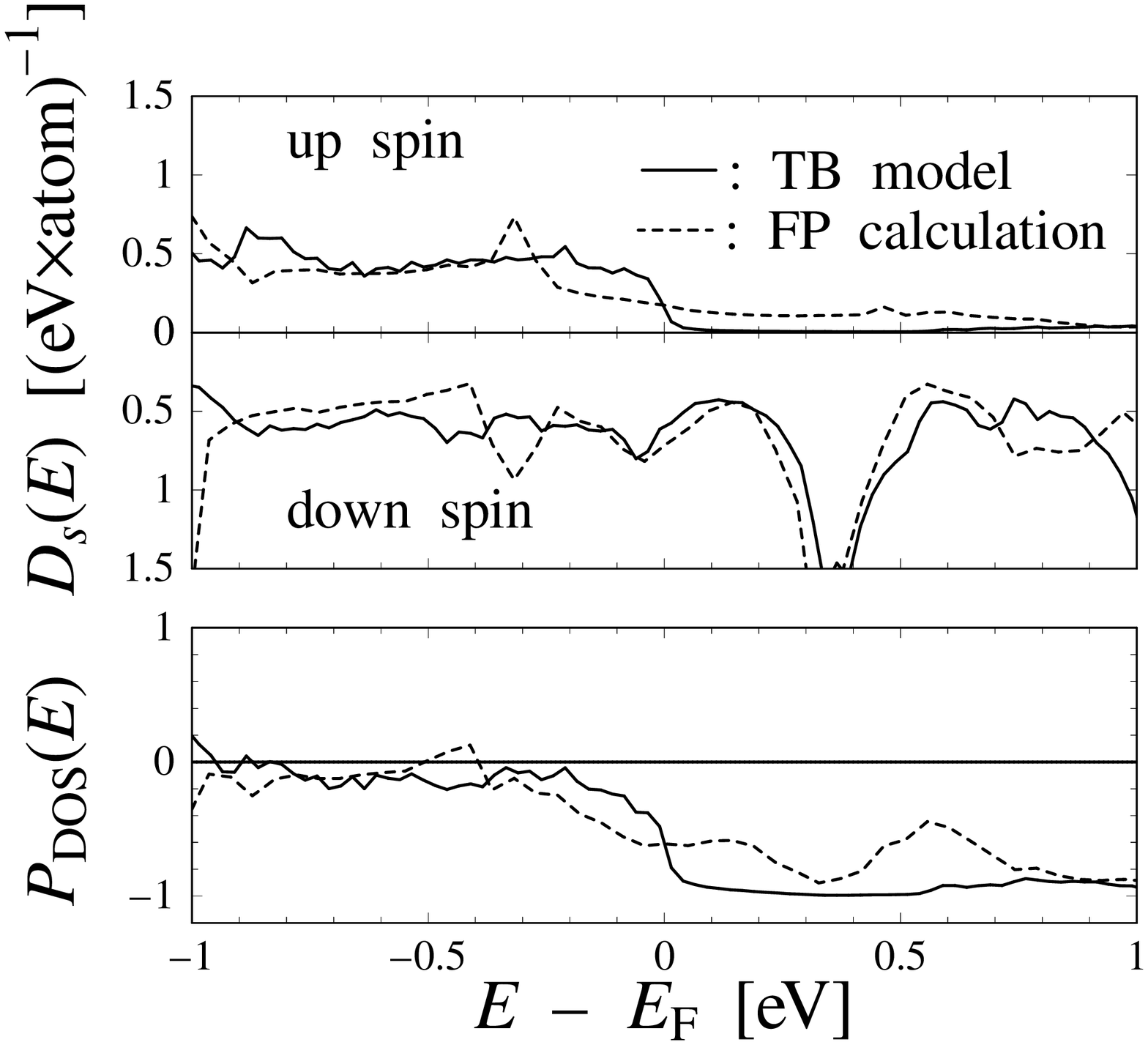}\\[-0.9cm]
\hspace*{-7.5cm}(c) \\
\vspace{0.3cm}
\includegraphics[width=0.73\linewidth]{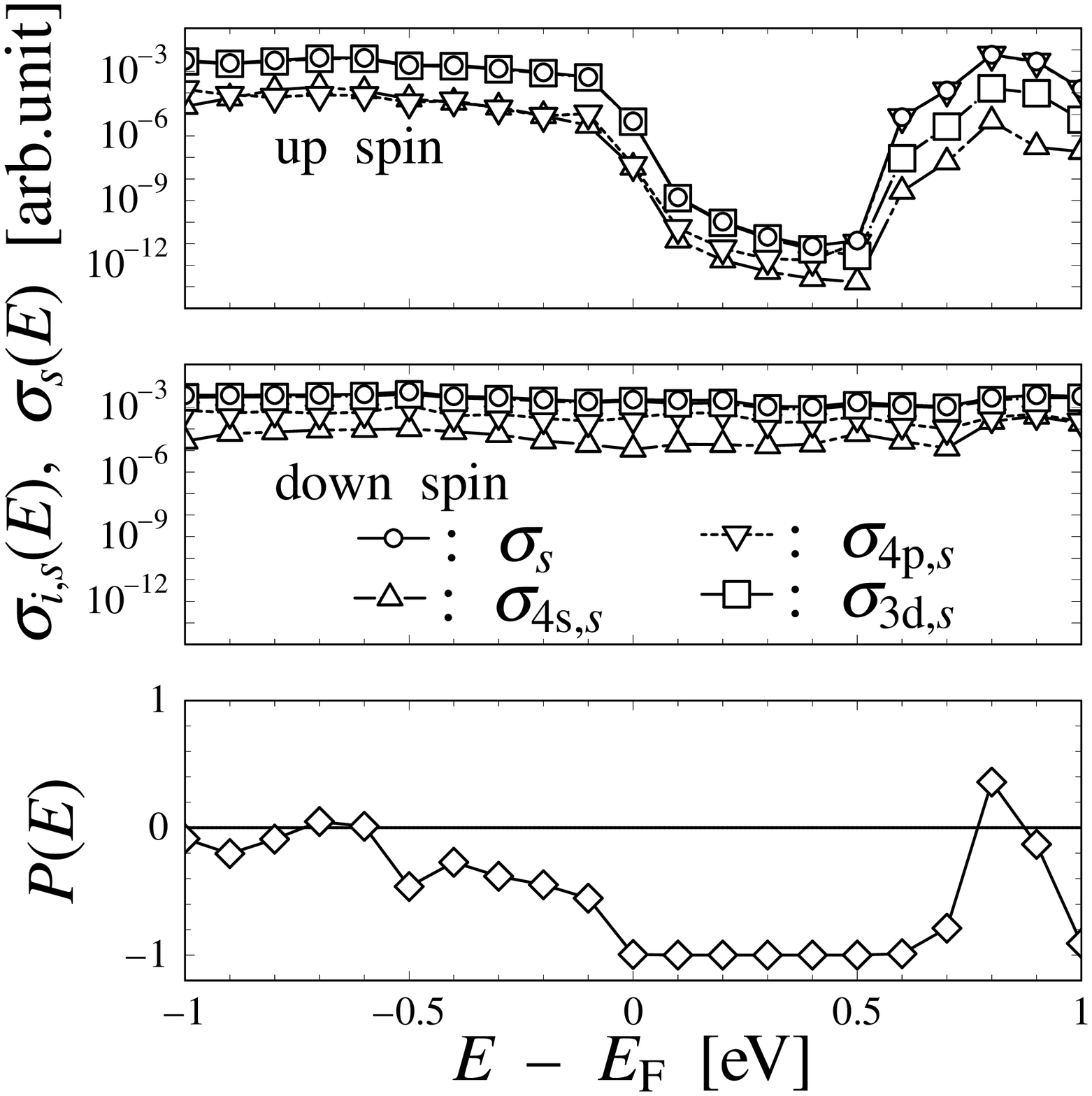}\\[-0.9cm]
\hspace*{-7.5cm}(d) \\
\caption{
Density of states, conductivities, and spin polarization ratios of Fe$_4$N: 
(a) $D_s(E)$ calculated using the FP calculation. 
(b) Schematic illustration of partial DOSs. 
The 4s-4p components are partly covered by the 3d ones 
[see darkish parts in 3d components]. 
(c) $D_s(E)$ and $P_{\mbox{\tiny DOS}}(E)$ for 
$-1~{\rm eV} \le E - E_{\mbox{\tiny F}} \le 1~{\rm eV}$ 
of the TB model and the FP calculation. 
(d) $\sigma_{i,s} (E)$, $\sigma_s (E)$, 
and $P(E)$ for $-1~{\rm eV} \le E - E_{\mbox{\tiny F}} \le 1~{\rm eV}$. 
Here, $\sigma_{\mbox{\tiny 2s},s}(E)$ 
and $\sigma_{\mbox{\tiny 2p},s}(E)$ for N 
are not shown because of their small values. 
}
\end{center}
\end{figure} 
\begin{figure}[ht]
\begin{center}
\includegraphics[width=0.73\linewidth]{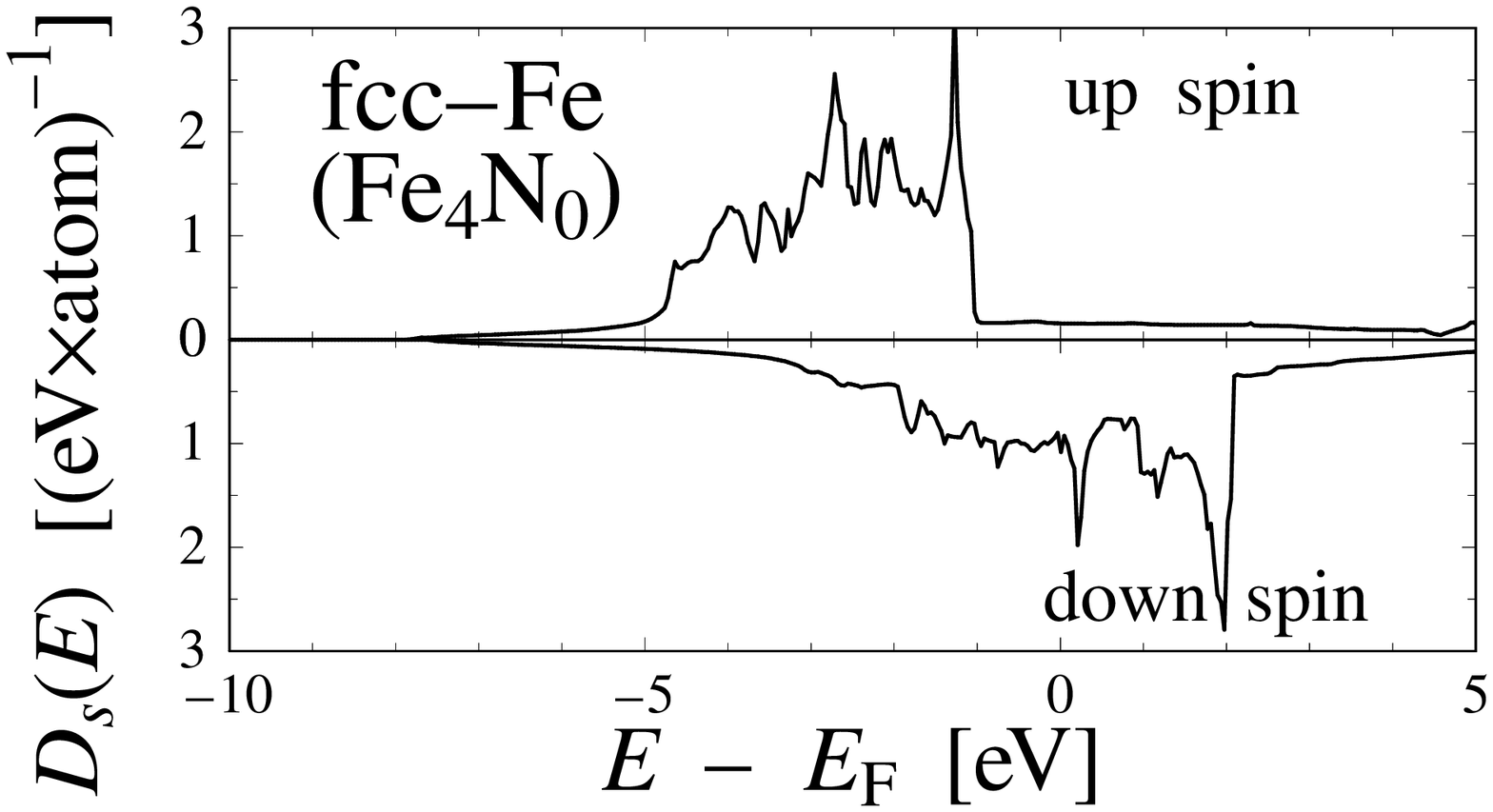}\\[-0.9cm]
\hspace*{-7.5cm}(a) \\
\vspace{0.6cm}
\hspace*{0.5cm}
\includegraphics[width=0.5\linewidth]{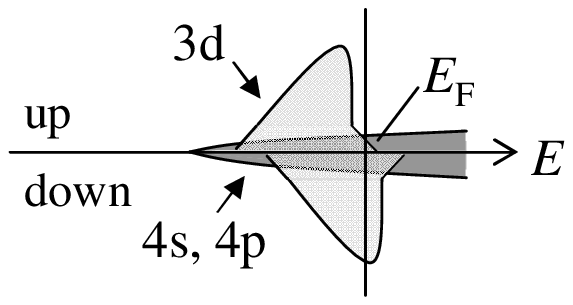}\\[-0.9cm]
\hspace*{-7.5cm}(b) \\
\vspace{0.2cm}
\includegraphics[width=0.73\linewidth]{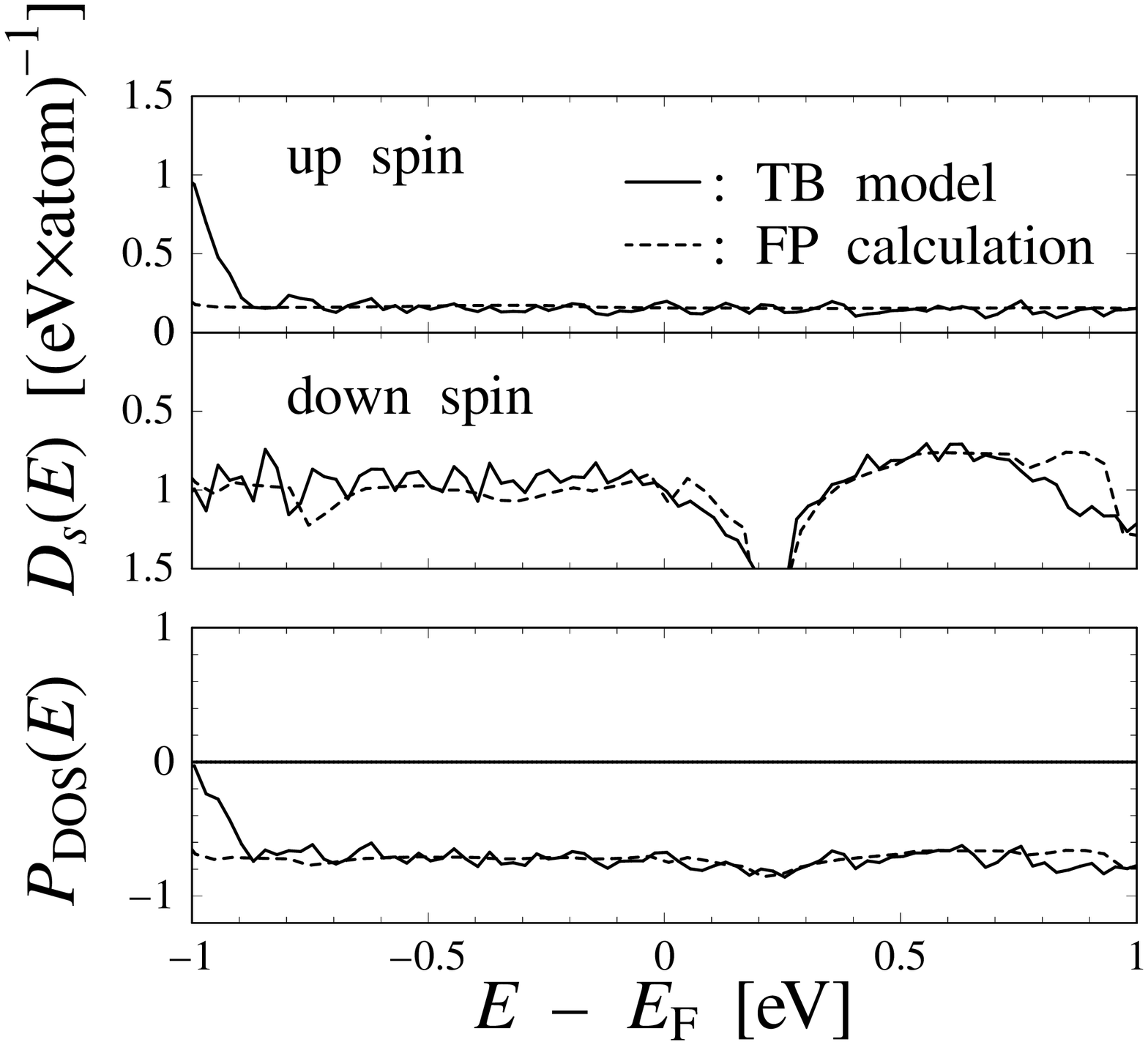}\\[-0.9cm]
\hspace*{-7.5cm}(c) \\
\vspace{0.3cm}
\includegraphics[width=0.73\linewidth]{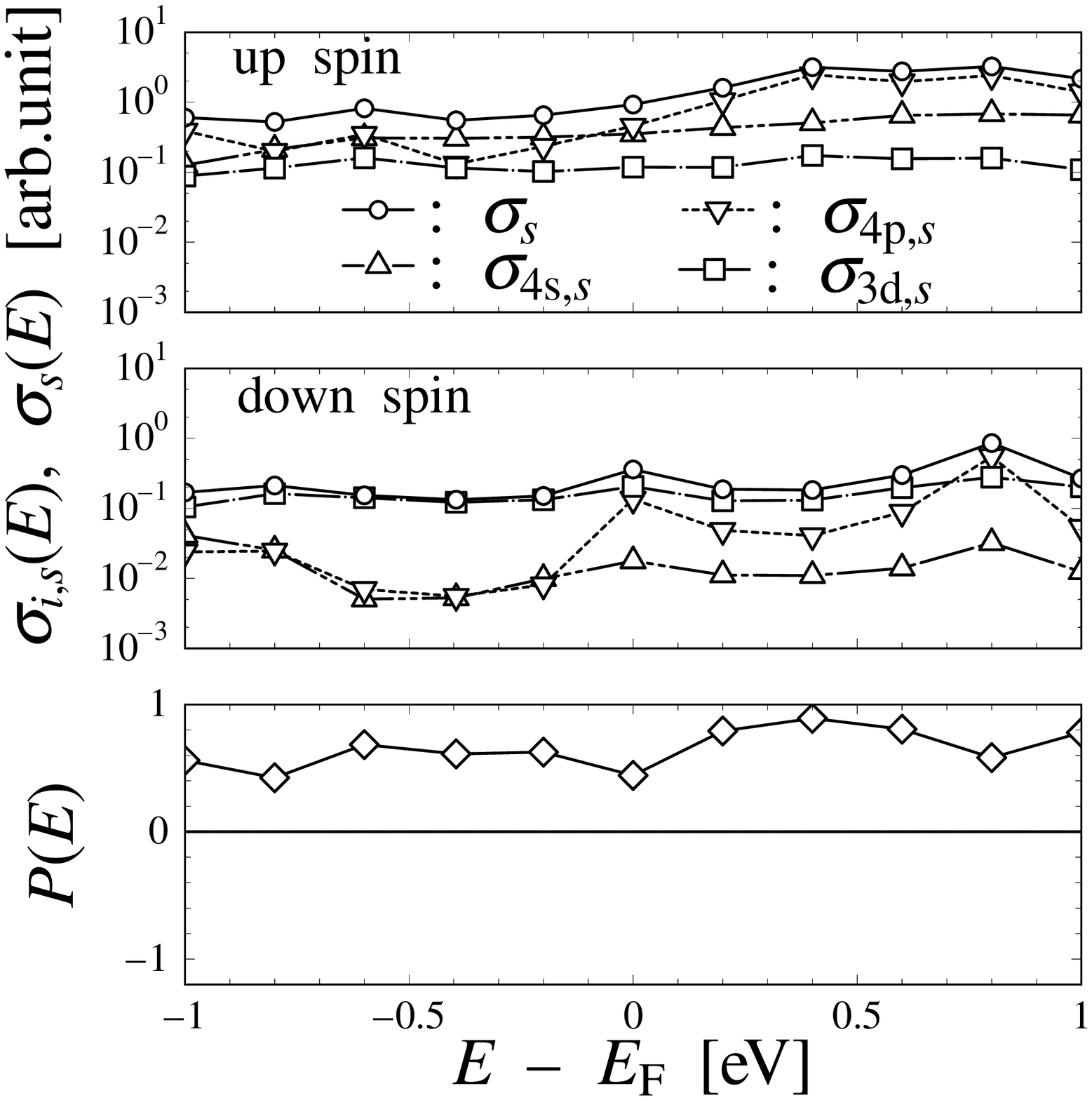}\\[-0.9cm]
\hspace*{-7.5cm}(d) \\
\caption{
The same as Fig. 1 for fcc-Fe (Fe$_4$N$_0$) 
in the ferromagnetic state with the equilibrium lattice constant of Fe$_4$N: 
(a) $D_s(E)$ calculated using the FP calculation. 
(b) Schematic illustration of partial DOSs. 
The 4s-4p components are partly covered by the 3d ones 
[see darkish parts in 3d components]. 
(c) $D_s(E)$ and $P_{\mbox{\tiny DOS}}(E)$ for 
$-1~{\rm eV} \le E - E_{\mbox{\tiny F}} \le 1~{\rm eV}$ 
of the TB model and the FP calculation. 
(d) $\sigma_{i,s} (E)$, $\sigma_s (E)$, 
and $P(E)$ for 
$-1~{\rm eV} \le E - E_{\mbox{\tiny F}} \le 1~{\rm eV}$. 
}
\end{center}
\end{figure}

In order to clarify the effect of an N atom on the transport, 
we investigated $D_s (E)$, $P_{\mbox{\tiny DOS}}(E)$, 
$\sigma_{i,s} (E)$, $\sigma_s (E)$, 
and $P(E)$ of fcc-Fe (Fe$_4$N$_0$) 
in the ferromagnetic state with an equilibrium lattice constant of Fe$_4$N. 
Regarding $D_s(E)$ around $E_{\mbox{\tiny F}}$ 
obtained using the FP calculation, 
$D_\downarrow (E)$ is much higher than $D_\uparrow (E)$ 
[see Figs. 2(a) and 2(c)]; 
for further details, 
low and broad DOSs of the 4s-4p orbitals of the up spin 
and a high DOS of the 3d orbitals of the down spin 
are observed [see Fig. 2(b)]. 
The SP ratio of DOS, 
$P_{\mbox{\tiny DOS}}(E_{\mbox{\tiny F}})$, 
is evaluated to be about $-$0.7. 
Figure 2(c) also shows $D_s (E)$ and $P_{\mbox{\tiny DOS}}(E)$ 
for $-1~{\rm eV} \le E \le 1~{\rm eV}$; 
these values are obtained using the TB model with the determined parameters. 
It is found that 
$D_s (E_{\mbox{\tiny F}})$ 
and $P_{\mbox{\tiny DOS}}(E_{\mbox{\tiny F}})$ of the TB model 
agree fairly well with the respective ones of the FP calculation. 
With the use of the TB model, 
we calculate $\sigma_{i,s} (E)$ and 
$\sigma_s (E)$ [see Fig. 2(d)]. 
In spite of the low DOSs of the 4s-4p orbitals of the up spin, 
their orbitals contribute strongly to $\sigma_\uparrow (E)$. 
This tendency originates from 
the large velocities of the 4s-4p orbitals. 
Moreover, $\sigma_{\mbox{\tiny 3d},\downarrow}(E)$ 
is relatively large 
reflecting the high DOS of the 3d orbitals of the down spin. 
The total conductivity at $E_{\mbox{\tiny F}}$ of the up spin 
$ \sigma_\uparrow (E_{\mbox{\tiny F}})$ 
then becomes larger than $\sigma_\downarrow (E_{\mbox{\tiny F}})$ 
even though $D_\uparrow (E_{\mbox{\tiny F}})$ is 
much lower than 
$D_\downarrow (E_{\mbox{\tiny F}})$. 
The SP ratio 
$P(E_{\mbox{\tiny F}})$ has a positive sign, 
opposite to the sign of $P_{\mbox{\tiny DOS}}(E_{\mbox{\tiny F}})$. 
This sign of $P(E_{\mbox{\tiny F}})$ is also 
opposite to that of Fe$_4$N. 
The magnitude of the SP ratio 
$|P(E_{\mbox{\tiny F}})|$ is evaluated to be 0.4, 
which is between the $|P(E_{\mbox{\tiny F}})|$ of Fe$_4$N and that of bcc-Fe.

On the basis of the above investigations 
and the evaluated Slater-Koster parameters, 
we discuss the role of the N atom 
on the highly spin-polarized transport of Fe$_4$N 
using schematic illustrations of partial DOSs 
[see Figs. 1(b) and 2(b)]. 
Note that 
by merely introducing the non-magnetic atom N to 
the body center position of fcc-Fe, 
$|P(E_{\mbox{\tiny F}})|$ becomes about 2.5 times 
as large as that of fcc-Fe 
and the sign of $P(E_{\mbox{\tiny F}})$ changes. 
In the present study, we find that, by adding N to fcc-Fe, 
4s-4p bands of Fe are raised to a higher energy region 
by a large magnitude of transfer integrals 
between the 4s-4p orbitals of Fe and the 2s and 2p orbitals of N, 
while 3d bands do not change significantly 
owing to the small magnitude of the transfer integrals 
between the 3d orbitals of Fe and the 2s and 2p orbitals of N. 
These behaviors are explained by considering 
the bonding-antibonding states formed by the transfer integrals, 
which correspond to overlaps between both orbitals. 
A large portion of 4s-4p bands exists 
in an energy region higher than 3d bands, 
and a small portion of them is located 
in the energy region of 3d bands 
because of hybridizations with 3d orbitals, 
and is then spin-polarized there. 
Hence, comparing Fe$_4$N with fcc-Fe, 
the proportion of 4s-4p bands is small 
in the vicinity of $E_{\mbox{\tiny F}}$, 
and, in particular, the proportion of the up spin is extremely small; 
the 3d bands of the down spin are dominant there. 
Therefore, the magnitude of the SP ratio $|P(E_{\mbox{\tiny F}})|$ increases, 
and $P(E_{\mbox{\tiny F}})$ becomes negative.

Finally, 
although it would be difficult to derive the general properties of phenomena 
from the present study alone, 
it seems possible that 
various ferromagnets consisting of magnetic elements and light elements 
have a large magnitude of SP ratios 
according to the same mechanism as that of Fe$_4$N. 
In addition, when such ferromagnets are used as the FMEs, 
FME/insulator/FME junctions may exhibit large MR ratios, 
although the MR ratios are often influenced by 
the interfacial states and materials of the insulator. 
In fact, CoFeB electrodes bring about a very large MR effect 
in CoFeB/MgO/CoFeB junctions.~\cite{Hayakawa1,Hayakawa2}

In conclusion, 
$|P(E_{\mbox{\tiny F}})|$ of Fe$_4$N was evaluated 
to be almost 1.0, 
which was about 5.0 times as large as that of bcc-Fe and 
about 2.5 times as large as that of fcc-Fe. 
In comparison with fcc-Fe, 
it was shown that the large magnitude of the SP ratio originated from 
the contribution to the transport of 3d bands, 
which was enhanced by introducing N. 
We anticipate that Fe$_4$N will become an electrode 
with a high efficiency of spin injection.
Furthermore, 
various ferromagnets consisting of magnetic elements and light elements 
may exhibit highly spin-polarized transport due to 
the present mechanism.

The authors thank Prof. T. Hoshino of Shizuoka University 
for useful discussions. 
One of the authors (S.K.) 
also thank members 
of nanomaterials theory group, AIST, for valuable discussions. 
This work has been supported 
by a competitive grant program 2005 of Shizuoka University.


\begin{thebibliography}{9}
\bibitem{bcc-Fe}
By performing the first principles calculation with the use of 
Vienna Ab-initio Simulation Package code, 
we evaluated $|P_{\mbox{\tiny DOS}} (E_{\mbox{\tiny F}})|$ of bcc-Fe 
to be about 0.3.
\end{thebibliography}
\end{document}